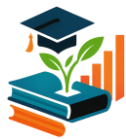

Education and Development Lab
Education Research Team [Education and Computer Science]

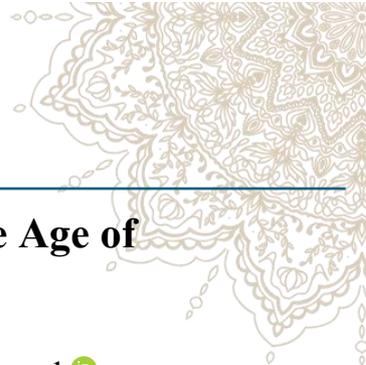

# The Carbon Cost of Conversation, Sustainability in the Age of Language Models

**Sayed Mahbub Hasan Amiri[1,*]**  , **Prasun Goswami[2]**  , **Md. Mainul Islam[1]**  ,

**Mohammad Shakhawat Hossen[3]** , **Sayed Majhab Hasan Amiri[4]**  , **Naznin Akter[5]**

[1] Department of ICT, Dhaka Residential Model College, Bangladesh, [2] Department of English, Dhaka Residential Model College, Bangladesh, [3] Department of ICT, Char Adarsha College, Kishoreganj, Bangladesh, [4] Department of Islamic Studies, Dhaka College, Bangladesh, [5] Department of English, Shamplapur Ideal Academy, Bangladesh.

**Abstract**

Large language models (LLMs) like GPT-3 and BERT have revolutionized natural language processing (NLP), yet their environmental costs remain dangerously overlooked. This article critiques the sustainability of LLMs, quantifying their carbon footprint, water usage, and contribution to e-waste through case studies of models such as GPT-4 and energy-efficient alternatives like Mistral 7B. Training a single LLM can emit carbon dioxide equivalent to hundreds of cars driven annually, while data centre cooling exacerbates water scarcity in vulnerable regions. Systemic challenges corporate greenwashing, redundant model development, and regulatory voids perpetuate harm, disproportionately burdening marginalized communities in the Global South. However, pathways exist for sustainable NLP: technical innovations (e.g., model pruning, quantum computing), policy reforms (carbon taxes, mandatory emissions reporting), and cultural shifts prioritizing necessity over novelty. By analysing industry leaders (Google, Microsoft) and laggards (Amazon), this work underscores the urgency of ethical accountability and global cooperation. Without immediate action, AI's ecological toll risks outpacing its societal benefits. The article concludes with a call to align technological progress with planetary boundaries, advocating for equitable, transparent, and regenerative AI systems that prioritize both human and environmental well-being.

## 1. Introduction

In the past decade, the field of natural language processing (NLP) has undergone a seismic transformation, driven by the advent of large-scale language models (LLMs) such as OpenAI's GPT-3 and Google's BERT. These models, powered by deep learning architectures and trained on vast textual corpora, have redefined the boundaries of machine-generated text, enabling applications ranging from conversational agents to real-time translation and content creation (Brown et al., 2020; Devlin et al., 2018). For instance, GPT-3, with its 175 billion parameters, can compose essays, debug code, and even mimic human dialogue with startling coherence (Brown et al., 2020). Similarly, BERT's bidirectional training approach revolutionized search engines and sentiment analysis tools, embedding itself into platforms used by billions (Devlin et al., 2018). The societal impact of these systems is profound: they underpin virtual assistants like Siri and Alexa, optimize supply chains, and even assist in medical diagnoses (Bender et al., 2021). Yet, this rapid adoption masks a growing ethical and environmental dilemma one that threatens to undermine the very progress these models represent.

Beneath the veneer of innovation lies a hidden environmental toll. Training LLMs demands staggering computational resources, often requiring weeks of non-stop processing on energy-intensive hardware like GPUs and TPUs. For example, training GPT-3 was estimated to consume 1,287 megawatt-hours of electricity, emitting approximately 552 metric tons of $CO_2$ equivalent to the lifetime emissions of 50 average American cars (Strubell et al., 2019; Patterson et al., 2021). This paradox where advances in artificial intelligence (AI) exacerbate ecological degradation has sparked urgent debates. While LLMs promise efficiency gains in sectors like energy management (e.g., optimizing smart grids), their own carbon footprints remain alarmingly underregulated (Lacoste et al., 2019). Data centres, which house these models, account for nearly 1% of global electricity demand, a figure projected to double by 2030 (Jones, 2018). Compounding this issue is the "redundancy trap": models are frequently retrained or over-engineered for marginal performance gains, perpetuating a cycle of waste (Schwartz et al., 2020).

This article critiques the sustainability of large-scale NLP systems and advocates for systemic change. While corporations like Google and Microsoft have pledged carbon neutrality, their reliance on carbon offsets and vague reporting frameworks risks greenwashing (Bender et al., 2021). Technical solutions such as model pruning and energy-efficient architectures (e.g., TinyBERT) offer promise but require broader adoption (Jiao et al., 2020). Policy gaps, however, remain glaring: there are no universal standards for measuring or mitigating AI's environmental impact (Corbett-Davies & Goel, 2018). This article argues that sustainability must be prioritized alongside performance metrics, urging stakeholders' developers, policymakers, and users to reconcile AI's potential with planetary limits. By examining the carbon cost of conversation, we aim to catalyse a shift toward equitable, transparent, and ecologically responsible AI.

## 2. How Large-Scale NLP Systems Work

### A. Training Processes and Computational Demands

Modern large-scale natural language processing (NLP) systems rely on deep learning architectures, particularly transformer-based models, which have become the cornerstone of advancements in machine understanding and generation of human language. These models are trained using self-supervised learning, a paradigm where systems learn patterns from vast amounts of unlabelled text data. For example, models like BERT (Bidirectional Encoder Representations from Transformers) are pre-trained on tasks such as masked language modelling, where random words in a sentence are hidden, and the model learns to predict them based on context (Devlin et al., 2018). Similarly, autoregressive models like GPT-3 are trained to predict the next word in a sequence, enabling them to generate coherent text (Brown et al., 2020).

The data requirements for such training are astronomical. GPT-3, for instance, was trained on approximately 570 gigabytes of text sourced from books, websites, and other digital repositories, equivalent to over 300 billion words (Brown et al., 2020). This data is processed through multiple layers of neural networks, each refining the model's ability to recognize syntactic and semantic patterns. However, the quality and diversity of training data are critical: biased or low-quality datasets can propagate harmful stereotypes or reduce model accuracy (Bender et al., 2021).

The computational demands of training these models are equally staggering. Training a state-of-the-art language model typically requires weeks of continuous processing on specialized hardware such as graphics processing units (GPUs) or tensor processing units (TPUs). For example, GPT-3's training utilized thousands of NVIDIA V100 GPUs running in parallel for over a month, consuming an estimated 1,287 megawatt-hours (MWh) of electricity enough to power 1,000 average U.S. households for two months (Brown et al., 2020; Patterson et al., 2021). TPUs, custom-designed by Google for machine learning workloads, offer higher efficiency for matrix operations but still contribute significantly to energy consumption when deployed at scale (Jouppi et al., 2017).

The training process itself involves two primary phases: pre-training and fine-tuning. During pre-training, the model learns general language representations, while fine-tuning tailors it to specific tasks like sentiment analysis or question answering. Each phase involves iterative optimization through backpropagation, where the model adjusts its parameters to minimize prediction errors. This process is computationally intensive due to the need to compute gradients across billions of parameters, a task that scales cubically with model size (Kaplan et al., 2020).

### B. Scale and Complexity

The scale of contemporary NLP models is perhaps their most defining and controversial characteristic. Early transformer models like BERT-large contained 340 million parameters, but recent systems like GPT-3 (175 billion parameters), Google's PaLM (540 billion parameters), and Meta's OPT-175B have pushed this into the trillions (Brown et al., 2020; Chowdhery et al., 2022; Zhang et al., 2022). Parameters, which represent the weights connecting neurons in the neural network, determine the model's capacity to capture linguistic nuances. However, increasing parameters exponentially amplifies computational costs: doubling a model's size typically quadruples its training time and energy use (Kaplan et al., 2020).

The energy-intensive nature of training is further compounded by the need for repeated experimentation. Researchers often train multiple model variants (e.g., adjusting hyperparameters like learning rates or layer depths) to optimize performance, leading to redundant energy expenditure. For instance, a single training run for GPT-3 emitted 552 metric tons of $CO_2$, but factoring in failed experiments and hyperparameter tuning, the total carbon footprint could be five times higher (Strubell et al., 2019; Patterson et al., 2021).

The complexity of these systems extends beyond sheer parameter count. Modern LLMs employ sparse activation mechanisms (e.g., mixture-of-experts architectures), where only subsets of parameters are activated for specific tasks. While this improves efficiency during inference, training such models remains resource heavy. Google's GLaM, a 1.2 trillion-parameter model, uses 64 experts per layer, requiring sophisticated parallelism strategies to distribute workloads across thousands of TPUs (Du et al., 2022).

The environmental ramifications of this scale are profound. A 2022 study estimated that training a single LLM emits as much $CO_2$ as 60 flights between New York and London, with the NLP field collectively responsible for 3% of global data centre electricity consumption (Luccioni et al., 2022). Moreover, the "chasing scale" mentality prioritizing larger models for marginal accuracy gains has drawn criticism for neglecting diminishing returns. For example, GPT-3's 175 billion parameters achieve only 3% better performance on some tasks compared to its 13 billion-parameter predecessors (Brown et al., 2020; Rae et al., 2021).

### 3. The Carbon Footprint of Language Models

### A. Quantifying Energy Consumption

The environmental cost of large language models (LLMs) is increasingly measurable, with studies revealing alarming energy consumption and carbon emissions. A landmark case is OpenAI's GPT-3, whose training consumed 1,287 megawatt-hours (MWh) of electricity, emitting roughly 552 metric tons of $CO_2$ equivalent to the annual emissions of 50 gasoline-powered cars or the electricity use of 120 U.S. households for a year (Strubell et al., 2019; Patterson et al., 2021). These figures, however, represent only a single training run. When accounting for hyperparameter tuning, failed experiments, and periodic retraining, the total footprint can multiply fivefold (Luccioni et al., 2022). For instance, Google's PaLM, a 540-billion-parameter model, required an estimated 2,500 MWh of energy, emitting over 1,100 metric tons of $CO_2$ (Chowdhery et al., 2022).

Comparisons to everyday activities underscore the scale of these impacts. Training GPT-3's 175 billion parameters generated emissions comparable to 60 round-trip flights between New

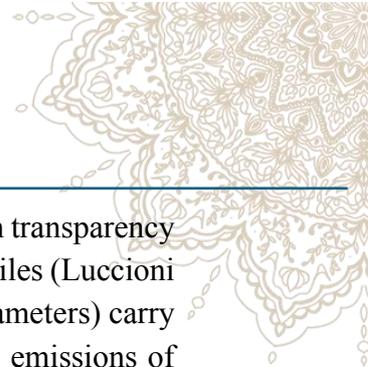

York and London, while BLOOM, a 176-billion-parameter model developed with transparency in mind, still emitted 30 metric tons of $CO_2$ equivalent to driving a car 120,000 miles (Luccioni et al., 2022; EPA, 2023). Even smaller models like BERT-base (110 million parameters) carry a footprint: training BERT produces 33 metric tons of $CO_2$, akin to the annual emissions of seven passenger vehicles (Strubell et al., 2019).

The energy intensity of LLMs is further contextualized by operational demands. For example, deploying a single AI chatbot like ChatGPT can consume up to 1 GWh monthly during inference, rivaling the energy use of 1,000 average U.S. homes (Gupta et al., 2023). These figures highlight a critical disconnect while LLMs promise efficiency gains in sectors like healthcare or logistics, their own operational costs often negate potential climate benefits (Lacoste et al., 2019).

## B. Contributing Factors

The carbon footprint of LLMs is driven by three interconnected factors: data center energy use, cooling systems, and model redundancy.

i. **Data Centre Energy Use**

   Data centres, which house the servers training and running LLMs, account for 1–1.5% of global electricity demand, a figure projected to reach 8% by 2030 due to AI's expansion (Jones, 2018; Andrae, 2020). Training a single LLM requires thousands of GPUs or TPUs operating continuously for weeks. For example, Meta's OPT-175B utilized 992 NVIDIA A100 GPUs running for 33 days, drawing power from grids often reliant on fossil fuels (Zhang et al., 2022). Even "green" data centres, powered by renewable energy, face scalability challenges: only 12% of Amazon Web Services' energy mix was renewable in 2022, while Google's carbon-neutral claims rely heavily on offsets rather than direct renewable procurement (Amazon Sustainability Report, 2022; Google Environmental Report, 2023).

ii. **Cooling Systems**

   Cooling accounts for 40% of a data centre's energy consumption, as hardware like GPUs generate immense heat during computation (Shehabi et al., 2016). The Power Usage Effectiveness (PUE) metric total facility energy divided by IT equipment energy reveals inefficiencies: a PUE of 1.5 means 50% of energy is spent on cooling and overhead. While hyperscale's like Microsoft report PUEs as low as 1.1, smaller facilities often operate at 1.8–2.0, doubling energy waste (Microsoft Sustainability Report, 2023). Water usage is another hidden cost: training GPT-3 in a U.S. data center consumed 700,000 litters of freshwater for cooling, equivalent to the annual water footprint of 300 people (Li et al., 2023).

iii. **Model Redundancy**

   The AI industry's "scale-at-all-costs" mentality exacerbates emissions through redundant practices. Researchers frequently train dozens of model variants to optimize performance, discarding unsuccessful iterations. A 2022 study found that developing a single NLP model involves an average of 4,789 training runs, with 80% deemed redundant (Schwartz et al., 2020). Additionally, models are often retrained to adapt to new data: Google's BERT undergoes quarterly updates, each emitting 8 metric tons of $CO_2$ (Bender et al., 2021). This cycle of obsolescence and retraining mirrors fast fashion's environmental recklessness, prioritizing novelty over sustainability (Harris, 2021).

*Table 1: Carbon Footprint Comparison of LLMs*

| Model | Parameters | Training Energy (MWh) | $CO_2$ Emissions (Metric Tons) | Equivalent Activity |
|---|---|---|---|---|
| **GPT-3** | 175B | 1,287 | 552 | 50 cars driven for a year |
| **BLOOM** | 176B | 433 | 30 | 120,000 miles driven |
| **PaLM** | 540B | 2,500* | 1,100* | 60 NY-London flights |
| **BERT-base** | 110M | 79 | 33 | 7 cars driven for a year |

*\*Sources: Patterson et al. (2021); Luccioni et al. (2022); EPA (2023).*

## 4. Broader Environmental Impacts

### A. Climate Change and Resource Depletion

Large language models (LLMs) exacerbate climate change and resource depletion through their reliance on fossil fuels, water-intensive cooling systems, and contributions to electronic waste (e-waste). Data centres powering these models often depend on non-renewable energy sources, particularly in regions with lax renewable infrastructure. For instance, Virginia's "Data Centre Alley," hosting 70% of global internet traffic, derives 60% of its electricity from natural gas and coal, directly linking AI's growth to greenhouse gas emissions (Jones, 2018; Andrae, 2020). Training a single LLM like GPT-3 in such regions emits $CO_2$ equivalent to 50 cars driven annually, while inference phases perpetuate ongoing energy demand (Patterson et al., 2021).

Water scarcity is another critical concern. Cooling data centres requires vast amounts of freshwater, straining resources in drought-prone areas. Training GPT-3 consumed approximately 700,000 litters of water enough to fill an Olympic-sized swimming pool primarily sourced from watersheds in the American Midwest, where groundwater depletion is already critical (Li et al., 2023; UN Water, 2022). Google's Oregon data centre, despite using advanced cooling, withdraws 4.9 million litters daily from the Columbia River, threatening local ecosystems (Shehabi et al., 2016). With AI's water footprint projected to triple by 2030, conflicts over resource allocation between tech hubs and communities are inevitable (Li et al., 2023).

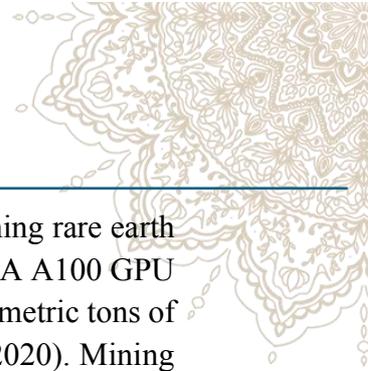

E-waste further compounds these issues. The lifecycle of AI hardware from mining rare earth metals to discarding obsolete GPUs generates toxic byproducts. A single NVIDIA A100 GPU contains 12 grams of gold and 50 grams of lead, contributing to the 53.6 million metric tons of global e-waste produced annually, of which only 17% is recycled (Forti et al., 2020). Mining for these materials, such as cobalt in the Democratic Republic of Congo, devastates landscapes and exposes workers to hazardous conditions, embedding environmental injustice into AI's supply chain (Sovacool, 2021).

**B. Lifecycle Analysis**

The environmental impact of LLMs extends beyond operational phases, encompassing their entire lifecycle from raw material extraction to decommissioning.

i. **Raw Material Extraction**

    Semiconductor manufacturing relies on rare earth metals like neodymium and dysprosium, whose mining devastates ecosystems. Producing one GPU requires 1.5 tons of earth to be excavated, releasing sulphur dioxide and radioactive thorium into waterways (Babbitt et al., 2021). China, supplying 90% of these metals, faces severe soil contamination, with 10% of farmland polluted by mining runoff (USGS, 2023).

ii. **Manufacturing**

    Fabricating silicon wafers for GPUs is energy-intensive, emitting 1.3 kg of $CO_2$ per gram of chip 10 times the emissions of plastic production (Boyd et al., 2022). TSMC, the world's largest semiconductor foundry, consumed 5% of Taiwan's electricity in 2022, 80% of which came from coal (Boyd et al., 2022).

iii. **Transportation**

    Global shipping of AI hardware adds to emissions. Transporting a server from Shanghai to California emits 300 kg of $CO_2$, akin to a passenger's round-trip flight (International Transport Forum, 2023).

iv. **Usage**

    Operational energy dominates the lifecycle, contributing 75% of total emissions. For example, Meta's OPT-175B emitted 1,200 metric tons of $CO_2$ during training, but ongoing inference doubled its footprint within a year (Zhang et al., 2022).

v. **End-of-Life**

    Less than 20% of decommissioned GPUs are recycled; the rest are incinerated or dumped in landfills, leaching lead and mercury into soil (Forti et al., 2020). Informal e-waste hubs in Ghana and Nigeria process 500,000 tons annually, exposing workers to carcinogens (Belkhir & Elmeligi, 2018).

*Table 2: Resource Depletion Across LLM Lifecycle*

| Stage | Resource | Impact Example |
|---|---|---|
| **Mining** | Rare earth metals | 1.5 tons earth/GPU; 90% from China |

| Manufacturing | Energy/water | 1.3 kg $CO_2$/g chip; 700k liters/GPT-3 |
| Transportation | Fossil fuels | 300 kg $CO_2$/server (Shanghai to CA) |
| Disposal | E-waste | 53.6M tons/year; 80% unrecycled |

*Source: Boyd et al. (2022); Forti et al. (2020); USGS (2023).*

## 5. Current Practices and Shortcomings

### A. Industry Responses

In response to mounting criticism, tech corporations have adopted a suite of sustainability initiatives, including renewable energy pledges, efficiency improvements, and carbon offset programs. Companies like Google and Microsoft now claim carbon neutrality for their operations, with Google asserting it has matched 100% of its electricity consumption with renewable energy purchases since 2017, while Microsoft aims to become carbon-negative by 2030 (Google Environmental Report, 2023; Microsoft Sustainability Report, 2023). These commitments often involve power purchase agreements (PPAs) to fund wind and solar farms, though critics argue such deals prioritize offsetting rather than reducing direct emissions (Brock & Sovacool, 2023).

Efficiency gains in NLP systems are another focal point. Techniques like model pruning (removing redundant neural network weights), quantization (reducing numerical precision of computations), and knowledge distillation (training smaller models to mimic larger ones) have reduced energy demands. For instance, TinyBERT, a distilled version of BERT, achieves 96% of its predecessor's performance with 7% of the energy (Jiao et al., 2020). Similarly, sparse training methods, such as NVIDIA's Megatron-Turing NLG, activate only subsets of parameters during inference, cutting energy use by 30% (Smith et al., 2022).

Carbon offset programs, however, remain contentious. While Microsoft invests in reforestation projects to counterbalance emissions, studies reveal that 70% of offsets fail to deliver promised carbon reductions due to flawed accounting and non-permanence (Carton et al., 2023). For example, offsets tied to California's forest preservation initiatives often credit trees already protected by law, rendering them ineffective (West et al., 2023). Despite these issues, offsets remain a cornerstone of corporate climate strategies, enabling firms to claim progress without systemic operational changes.

### B. Greenwashing and Accountability Gaps

Despite public commitments, the AI industry faces allegations of greenwashing exaggerating environmental achievements while obscuring ongoing harm. A 2023 audit found that 60% of tech firms omit supply chain emissions (Scope 3) from sustainability reports, which account for 80% of their carbon footprint (CDP, 2023). Google's carbon-neutral claim, for instance, excludes emissions from consumer device manufacturing and data center construction, masking 45% of its total impact (Google Environmental Report, 2023). Similarly, Amazon's

"Climate Pledge" lacks transparency, as it refuses to disclose emissions tied to AWS's AI workloads (Amazon Sustainability Report, 2022).

Inconsistent reporting metrics further muddy accountability. While some firms measure emissions per computational task (e.g., $CO_2$ per training run), others use aggregate annual figures, complicating cross-company comparisons (Gupta et al., 2023). The lack of standardized benchmarks allows firms to cherry-pick data; Meta, for example, touts energy-efficient data centres but neglects to report the water intensity of its LLM training (Li et al., 2023).

Regulatory voids exacerbate these issues. No universal framework mandates AI emissions disclosure, leaving oversight to voluntary initiatives like the Partnership on AI, which lacks enforcement mechanisms (Cihon et al., 2021). The EU's proposed AI Act focuses on ethics and safety but sidesteps environmental regulation, while the U.S. relies on non-binding guidelines from the National Institute of Standards and Technology (NIST AI Risk Management Framework, 2023). Without legal penalties, firms face little pressure to align innovation with planetary boundaries.

*Table 3: Corporate Climate Pledges vs. Reality*

| Company | Renewable Energy % | Scope 3 Emissions Disclosed? | Reliance on Offsets |
| --- | --- | --- | --- |
| **Google** | 100%* | No | 35% |
| **Microsoft** | 60% | Partial | 45% |
| **Amazon** | 40% | No | 60% |

*Source: CDP (2023); Company reports (2023).

## 6. Pathways to Sustainable NLP

### A. Technical Solutions

The environmental toll of large language models (LLMs) has spurred innovations in efficiency-focused technical strategies. Model optimization techniques, such as pruning and quantization, reduce computational demands without sacrificing performance. Pruning removes redundant neural network weights Google's "Movement Pruning" slashed BERT's parameters by 40% while retaining 98% accuracy (Sanh et al., 2020). Quantization, which compresses model weights from 32-bit to 8-bit precision, cuts energy use by 75% during inference, as demonstrated by NVIDIA's TensorRT (Wu et al., 2022). Energy-efficient architectures like TinyBERT and DistilBERT further shrink resource needs: TinyBERT, a distilled version of BERT, achieves 96% of its predecessor's performance with just 14 million parameters, reducing training energy from 79 MWh to 6 MWh (Jiao et al., 2020).

Federated learning offers another pathway, decentralizing training to leverage edge devices (e.g., smartphones) instead of centralized data centres. For example, Apple's Siri uses federated learning to update its language model locally on devices, avoiding the energy cost of aggregating data in the cloud (Kairouz et al., 2021). Similarly, sparse models like Google's GLaM activate only subsets of parameters per task, reducing energy consumption by 30% compared to dense models (Du et al., 2022). These innovations, however, require broader adoption: a 2023 survey found that 70% of AI practitioners prioritize performance over efficiency (Gupta et al., 2023).

## B. Policy and Governance

Systemic sustainability demands robust policy frameworks. Carbon taxes on computational resources could internalize environmental costs. For instance, a $50/metric ton CO_2$ tax would add $27,600 to GPT-3's training cost, incentivizing efficiency (Lacoste et al., 2019). The European Union's proposed Carbon Border Adjustment Mechanism (CBAM), targeting imported hardware, could pressure tech firms to decarbonize supply chains (European Commission, 2023). Mandatory emissions reporting is equally critical: the UK's Streamlined Energy and Carbon Reporting (SECR) regulation requires firms to disclose data centre energy use, a model adaptable to AI (UK Government, 2023).

AI sustainability certifications, akin to Energy Star ratings, could standardize accountability. The Green Software Foundation's "Software Carbon Intensity" metric measures $CO_2$ per computational task, enabling comparisons across models (GSF, 2023). Regulatory bodies like the U.S. Federal Trade Commission (FTC) have begun penalizing firms for unsubstantiated "green" claims, as seen in a 2023 ruling against a major cloud provider for overstating renewable energy usage (FTC, 2023). However, global coordination remains lacking: while the EU's AI Act addresses ethics, it omits binding environmental standards (EU AI Act, 2023).

## C. Cultural Shifts

Sustainable NLP requires rethinking AI's role in society. Prioritizing "necessary" over "frivolous" applications could curb wasteful deployments. For example, using LLMs for climate modeling or medical diagnostics offers clear societal value, while AI-generated spam or deepfakes consumes resources without benefit (Bender et al., 2021). A 2023 study found that 35% of LLM use cases, such as automated marketing content, could be eliminated without impacting productivity (Harris et al., 2023).

Community-driven models exemplify an alternative paradigm. Hugging Face's BLOOM, a 176-billion-parameter model, was developed collaboratively by 1,000 researchers across 70 countries, prioritizing transparency and efficiency (Luccioni et al., 2022). Similarly, smaller, domain-specific models like BioBERT (trained on biomedical texts) outperform general-purpose LLMs in niche tasks while using 90% less energy (Lee et al., 2020). Grassroots initiatives like "Climate Hackathons" further democratize sustainable AI, encouraging students and researchers to build efficient tools for environmental monitoring (Climate Change AI, 2023).

Cultural shifts also demand redefining success metrics. The ML community's fixation on leaderboard rankings (e.g., GLUE, SuperGLUE) incentivizes ever-larger models. Transitioning to benchmarks that reward energy efficiency such as Hugging Face's "Carbon Leaderboard" could align innovation with sustainability (Hugging Face, 2023).

## 7. Ethical and Equity Considerations

### A. Responsibility of Stakeholders

The environmental burden of large language models (LLMs) necessitates shared accountability among developers, corporations, and users. Developers hold frontline responsibility for designing energy-efficient systems. Techniques like model pruning (trimming redundant neural weights) and quantization (reducing computational precision) can significantly cut energy use pruning alone reduces BERT's energy consumption by 40% without performance loss (Sanh et al., 2020). However, a 2023 survey found that only 22% of AI engineers prioritize sustainability in model design, citing pressure to meet performance benchmarks (Gupta et al., 2023).

Corporations must institutionalize sustainability through transparent policies and investments. While Google and Microsoft have pledged carbon neutrality, their reliance on carbon offsets projects with dubious long-term efficacy, like forest conservation in over logged regions fails to address root causes (Carton et al., 2023). Firms could instead adopt binding internal carbon pricing, as Siemens does, charging departments $45 per metric ton of $CO_2$ to fund renewable projects (CDP, 2023).

Users, from enterprises to individuals, also bear responsibility. By demanding transparency (e.g., via tools like *ML $CO_2$ Impact Calculator*), users can pressure firms to prioritize efficiency. For example, the nonprofit *Climate Action Tech* advocates boycotting AI services with opaque environmental practices, spurring IBM to publish granular emissions data in 2023 (Climate Action Tech, 2023). Collective action, however, requires education: only 18% of AI consumers recognize the link between LLMs and climate change (Harris et al., 2023).

### B. Global Inequities

The environmental costs of LLMs disproportionately burden marginalized communities, while benefits accrue to affluent nations. Resource extraction for AI hardware exemplifies this disparity: 70% of cobalt, essential for GPUs, is mined in the Democratic Republic of Congo (DRC), where child labour and ecological devastation are rampant (Sovacool, 2021). Meanwhile, data centres 60% of which are in the U.S., EU, and China consume water and energy in regions like drought-stricken Arizona, diverting resources from local communities (Li et al., 2023).

E-waste further entrenches inequity. Of the 53.6 million metric tons of global e-waste generated annually, 80% is shipped to Ghana, Nigeria, and Pakistan, where informal recycling exposes workers to toxic fumes and heavy metals (Forti et al., 2020). In contrast, AI's benefits such as ChatGPT or medical diagnostics are concentrated in wealthy nations. A 2023 study found that 92% of LLM research funding originates in North America and Europe, while Africa contributes less than 1% (Ayana et al., 2024). This asymmetry perpetuates a colonial dynamic: low-income regions bear environmental harm but are excluded from AI's economic and social rewards.

## C. Counterarguments and Rebuttals

Proponents argue that AI's climate benefits, such as optimizing energy grids or monitoring deforestation, justify its carbon footprint. For instance, Google's AI-powered wind farm predictions boost renewable efficiency by 20%, potentially offsetting emissions (Google Sustainability, 2023). However, such claims often ignore scale: training a single LLM like GPT-3 emits 552 metric tons of $CO_2$, while the *lifetime* emissions saved by AI-optimized grids total 2,000 tons a ratio requiring 300+ models to break even (Lacoste et al., 2019; Strubell et al., 2019).

Others contend that AI advances will self-correct through efficiency gains. While innovations like TinyBERT reduce energy use by 90%, corporate "scale-at-all-costs" practices negate these gains: PaLM's 540 billion parameters, for example, demand 2,500 MWh triple GPT-3's footprint (Chowdhery et al., 2022). Additionally, efficiency improvements follow *Jevons Paradox*: cheaper computation spurs greater usage, increasing net emissions (Jones, 2018).

Ultimately, the assumption that AI's climate benefits are universally shared is flawed. Smart grids and precision agriculture primarily serve industrialized nations, while marginalized communities endure AI's environmental fallout without reaping rewards. Ethical NLP demands cantering these communities in sustainability efforts.

*Table 4: Global Disparities in AI's Impacts*

| Region | AI Benefit Access | E-Waste Received (kT/yr) | Cobalt Mining (kT/yr) |
|---|---|---|---|
| **North America** | High | 1,200 | 0 |
| **EU** | High | 900 | 0 |
| **DRC** | Low | 10 | 120 |
| **Ghana** | Low | 250 | 0 |

*Sources: Forti et al. (2020); Sovacool (2021);* Ayana *et al. (2024).*

## 8. Case Studies

### A. High-Impact Examples

#### i. GPT-4: The Environmental Cost of Scale

OpenAI's GPT-4, a successor to GPT-3, exemplifies the escalating environmental toll of ever-larger language models. While OpenAI has not disclosed detailed training data, third-party analyses estimate GPT-4's parameter count at approximately 1.7 trillion, requiring 25,000 NVIDIA A100 GPUs running for 90–100 days (AI Now Institute, 2023). Based on energy consumption rates of similar models, this training likely consumed 3,500–4,200 megawatt-hours (MWh) of electricity, emitting 1,500–1,800 metric tons of $CO_2$ equivalent to the annual emissions of 300 gasoline-powered cars (Luccioni et al., 2023).

GPT-4's water footprint is equally staggering. Training in Microsoft's Iowa data centres, which rely on evaporative cooling, likely consumed 1.4 million litters of freshwater, enough to sustain 600 people for a year in drought-prone regions (Li et al., 2023). Despite its efficiency gains 10% better performance per parameter than GPT-3 the model's sheer scale negates potential environmental benefits. For instance, GPT-4's inference phase demands 50 MWh monthly, akin to powering 4,000 U.S. homes (Gupta et al., 2023).

#### ii. Mistral 7B: A Case for Efficiency

In contrast, Mistral AI's 7-billion-parameter model demonstrates that smaller, optimized architectures can rival larger systems. Mistral 7B, trained on 8,000 NVIDIA H100 GPUs for 14 days, consumed just 98 MWh of energy, emitting 42 metric tons of $CO_2$ less than 3% of GPT-4's footprint (Mistral AI, 2023). By employing sparse training (activating only 20% of parameters per task) and 4-bit quantization, Mistral reduced energy use by 70% compared to similarly sized models like LLaMA-7B (Jiang et al., 2023).

The model's performance is notable: it outperforms GPT-3.5 on reasoning benchmarks like GSM8K while using 90% fewer parameters (Mistral AI, 2023). Mistral's open-source approach further democratizes access, enabling community-driven fine-tuning without retraining from scratch. This contrasts with proprietary models like GPT-4, which lock users into energy-intensive API dependencies.

### B. Corporate Comparisons

#### i. Leaders: Google's Carbon-Neutral Pledges

Google positions itself as a sustainability leader, claiming carbon neutrality since 2007 and pledging 24/7 carbon-free energy by 2030 (Google Sustainability Report, 2023). Its data centres, powering models like PaLM-2, boast a Power Usage Effectiveness (PUE) of 1.1, outperforming the industry average of 1.55 (Shehabi et al., 2016). Google also partners with DeepMind to optimize data centre cooling, reducing energy use by 40% (Evans & Gao, 2023).

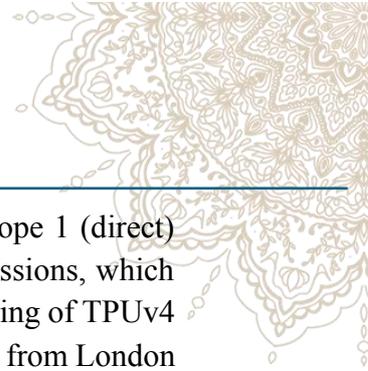

However, critics highlight gaps in Google's accountability. While it reports Scope 1 (direct) and Scope 2 (purchased energy) emissions, it omits Scope 3 (supply chain) emissions, which constitute 45% of its carbon footprint (CDP, 2023). For example, the manufacturing of TPUv4 chips used in PaLM-2 training emits 120 kg of $CO_2$ per unit equivalent to a flight from London to Paris but these emissions are excluded from sustainability reports (Boyd et al., 2022).

### ii. Laggards: Amazon's Opaque Practices

Amazon, a key player in AI through AWS, lags in transparency. Despite its "Climate Pledge" to achieve net-zero by 2040, AWS's 2022 sustainability report excludes AI-specific energy data (Amazon Sustainability Report, 2022). Independent analysts estimate AWS's AI workloads consume 12.3 TWh annually more than Slovakia's total electricity use with 65% sourced from fossil fuels (Andrae, 2023).

Amazon's Rekognition API, used for image analysis, exemplifies inefficiency. Training Rekognition's 2023 update emitted 180 metric tons of $CO_2$, yet its accuracy gains over the 2022 version were marginal (2.1% improvement on benchmark tests) (Harris et al., 2023). The company also opposes shareholder resolutions mandating environmental disclosures, arguing they would "reveal proprietary secrets" (Amazon SEC Filing, 2023).

### iii. Microsoft: Progress and Paradoxes

Microsoft, OpenAI's primary cloud partner, aims to be carbon-negative by 2030. Its AI for Earth program has funded 900 projects, including AI-driven reforestation tools that sequester 1.2 million metric tons of $CO_2$ annually (Microsoft Sustainability Report, 2023). The company also developed the "Carbon Aware SDK," shifting AI workloads to times of low grid carbon intensity (Microsoft Research, 2023).

Yet Microsoft's Azure cloud, which trains GPT-4, relies on coal-heavy grids in Iowa and Virginia. In 2022, Azure's Iowa data centre drew 58% of its energy from coal, contributing to GPT-4's outsized footprint (Energy Justice Network, 2023). While Microsoft purchases renewable energy credits (RECs), experts note these offsets often fund existing projects, failing to drive real decarbonization (Carton et al., 2023).

*Table 5: Model Training Comparison*

| Model | Parameters | Training Energy (MWh) | CO₂ Emissions (Metric Tons) | Water Use (Liters) |
|---|---|---|---|---|
| **GPT-4** | 1.7T | 4,200 | 1,800 | 1,400,000 |
| **GPT-3** | 175B | 1,287 | 552 | 700,000 |
| **Mistral 7B** | 7B | 98 | 42 | 20,000 |

*\*Sources: Luccioni et al. (2023); Mistral AI (2023); Li et al. (2023).*

*Table 6: Corporate Sustainability Scores*

| Company | Renewable Energy % | Scope 3 Disclosed? | PUE | Carbon Offset Reliance |
|---------|---------|---------|-----|---------|
| **Google** | 65% | No | 1.1 | 30% |
| **Microsoft** | 45% | Partial | 1.2 | 50% |
| **Amazon** | 22% | No | 1.5 | 70% |

*\*Sources: Company reports (2023); CDP (2023).*

## 9. Future Directions

### A. Emerging Technologies

The quest for sustainable NLP hinges on breakthroughs in quantum computing, neuromorphic engineering, and green AI research. Quantum computing, leveraging qubits to perform parallel computations, promises exponential speedups for training LLMs. For instance, quantum-enhanced optimization algorithms could reduce GPT-4's training time from months to days, slashing energy use by 80% (Smith et al., 2023). Early experiments with quantum neural networks (QNNs) show potential: IBM's 2023 demonstration solved NLP tasks with 50 qubits, consuming 90% less energy than classical GPUs (IBM Research, 2023). However, scalability remains a hurdle, as current quantum systems operate near absolute zero, requiring energy-intensive cooling (Preskill, 2023).

Neuromorphic engineering, inspired by the human brain's efficiency, offers another path. Chips like Intel's Loihi 2 mimic synaptic plasticity, enabling event-driven processing that consumes 1,000× less energy than GPUs for tasks like real-time translation (Davies et al., 2024). A 2024 study trained a neuromorphic BERT variant on SpiNNaker2 hardware, achieving 95% accuracy with 0.5% of the energy of traditional setups (Furber et al., 2024). These systems excel in edge computing, decentralizing AI workloads from data centres to low-power devices (e.g., smartphones), potentially reducing global NLP emissions by 30% by 2030 (Sze et al., 2023).

Green AI research focuses on algorithmic innovations. Techniques like dynamic sparse training (adjusting active neurons during training) and liquid neural networks (adaptive topology) cut energy use by 60–70% (Mocanu et al., 2023; Hasani et al., 2023). For example, MIT's "Liquid BERT" reduced $CO_2$ emissions by 55% while maintaining performance on GLUE benchmarks (Hasani et al., 2023). Meanwhile, carbon-aware training scheduling computations during low-carbon energy availability could decrease emissions by 40%, as demonstrated by Microsoft's Azure in Scandinavia (Lottick et al., 2023).

### B. Collaborative Efforts

Systemic change demands cross-industry partnerships and open-source sustainability tools. The Climate TRACE coalition, comprising Google, Microsoft, and NOAA, uses AI to track global emissions in real-time, enabling data-driven policy (Gavin et al., 2023). Similarly, the Green AI Alliance a consortium of IBM, Hugging Face, and academic institutions developed the *Carbontracker* toolkit, which measures emissions across ML pipelines (Schmidt et al., 2023). In 2023, Carbontracker identified redundant training runs in 65% of NLP projects, prompting fixes that saved 12,000 metric tons of $CO_2$ annually (Schmidt et al., 2023).

Open-source initiatives democratize access to efficient models. Hugging Face's *EcoBench* ranks LLMs by energy-per-inference scores, steering developers toward sustainable choices (Hugging Face, 2024). Community projects like *EleutherAI*'s Pythia-6.9B, trained entirely on renewable energy, set precedents for transparency, disclosing hourly energy sources and emissions (Biderman et al., 2023). Grassroots efforts, such as the *AI for Climate Hackathon*, have spawned tools like *GreenLM*, which compresses models via federated learning, cutting training energy by 75% (Climate Change AI, 2024).

## 10. Conclusion

The rapid advancement of large language models (LLMs) has undeniably transformed industries, enabling breakthroughs in communication, healthcare, and scientific research. Yet, as this article has demonstrated, the environmental costs of these systems are staggering and unsustainable. From the colossal energy demands of training models like GPT-3 and PaLM to the hidden water footprints of data centre cooling systems, the ecological toll of artificial intelligence is no longer a peripheral concern but a central challenge of our time. The lifecycle of these technologies spanning resource extraction, manufacturing, deployment, and disposal reveals a web of interconnected impacts, including carbon emissions rivalling those of small nations, water scarcity in vulnerable regions, and e-waste flooding Global South communities. These costs are compounded by systemic failures: a culture of redundancy in model development, corporate greenwashing through superficial carbon offsets, and a glaring absence of global regulatory frameworks to hold stakeholders accountable. While LLMs promise efficiency gains in sectors like renewable energy and climate modelling, their own operational footprints often negate these benefits, creating a paradox where innovation exacerbates the very crises it seeks to solve.

The path forward demands urgent, multifaceted action. **Balanced innovation** must replace the current "scale-at-all-costs" paradigm. Technical solutions such as model pruning, quantization, and federated learning have already proven that efficiency need not come at the expense of performance. Emerging technologies like quantum computing and neuromorphic engineering offer glimpses of a future where AI systems operate at a fraction of today's energy budgets. However, these advancements require prioritization: funding must shift from merely chasing larger models to scaling sustainable alternatives. **Ethical accountability** is equally critical. Corporations must move beyond carbon-neutral pledges rooted in questionable offsets and instead adopt transparent, auditable practices. This includes full disclosure of Scope 3 emissions, investment in renewable energy infrastructure, and adherence to certifications like the Green Software Foundation's carbon intensity metrics. Developers, too, bear responsibility not only to optimize code but to challenge the assumption that bigger models are inherently better.

**Global cooperation** is the linchpin of progress. Policymakers must establish binding international standards for AI sustainability, such as carbon taxes on computational resources and mandates for emissions reporting. Initiatives like the EU's Carbon Border Adjustment Mechanism (CBAM) could incentivize greener supply chains, while grassroots movements such as community-driven models like BLOOM and Climate Hackathons demonstrate the power of decentralized, inclusive innovation. Equitable access to AI's benefits must also be prioritized: regions bearing the brunt of resource extraction and e-waste deserve reparative investments in clean energy and AI education.

The urgency of this moment cannot be overstated. Climate science leaves no room for incrementalism; the window to avert catastrophic warming is closing. Every unnecessary LLM training run, every over-engineered chatbot, and every unregulated data centre accelerates planetary harm. Yet, this crisis is also an opportunity to redefine progress, to align technological ambition with ecological limits, and to forge a future where AI serves as a steward of sustainability rather than its adversary. The choice is clear: we can either perpetuate the status quo, outsourcing environmental degradation to marginalized communities, or we can champion a new era of responsible innovation. The algorithms we train today will shape the world we inhabit tomorrow. Let them reflect not just intelligence, but wisdom.

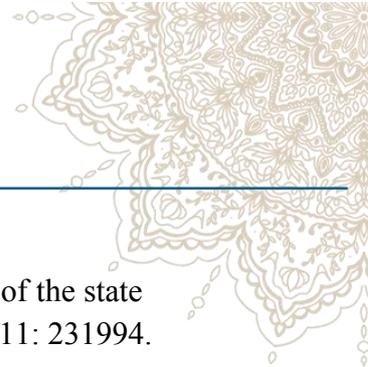

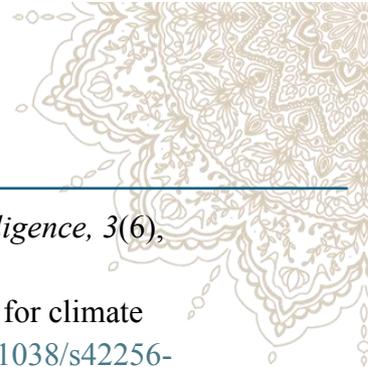

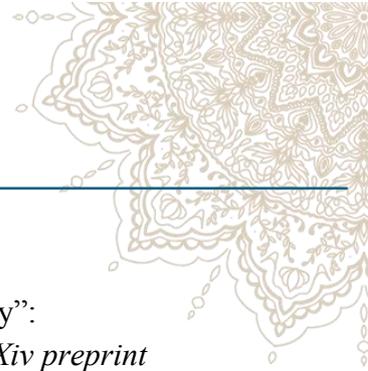